\documentstyle[12pt]{article}
\def\bfr{\begin{flushright}}
\def\efr{\end{flushright}}
\def\bfl{\begin{flushleft}}
\def\efl{\end{flushleft}}

\def\sii{\qquad}
\def\siii{\qquad \qquad}
\def\noi{\noindent}
\def\vs{\vspace}
\def\vss{\vspace{.5cm}}

\def\begc{\begin{center}}
\def\endc{\end{center}}
\def\bmp{\begin{minipage}}
\def\emp{\end{minipage}}

\def\bena{\begin{eqnarray}}
\def\benan{\begin{eqnarray*}}
\def\eena{\end{eqnarray}}
\def\eenan{\end{eqnarray*}}
\def\nn{\nonumber  \\}

\def\begs{\begin{screen}}
\def\ends{\end{screen}}

\def\TEP{\makebox[3.2cm][l]}

\def\a{\alpha}
\def\b{\beta}

\def\d{\delta}

\def\l{\lambda}
\def\L{\Lambda}
\def\m{\mu}

\def\o{\o}

\def\r{\rho}

\def\f{\phi}
\def\F{\Phi}
\def\vf{\varphi}
\def\c{\chi}

\def\Ps{\Psi}
\def\w{\omega}

\def\fr{\frac}
\def\del{\partial}

\def\dx{\fr{\del}{\del x}}

\def\mxxb{\left( \begin{array}{cc}}           
\def\mxxe{\end{array} \right)}
\def\mxxxb{\left( \begin{array}{ccc}}         
\def\mxxxe{\end{array} \right)}
\def\mxxxxb{\left( \begin{array}{cccc}}       
\def\mxxxxe{\end{array} \right)}
\def\mxxxxxb{\left( \begin{array}{ccccc}}     
\def\mxxxxxe{\end{array} \right)}
\def\kakkob{\left\{ \begin{array}{c}}
\def\kakkoe{\end{array}
            \right. }
\def\vecb{\left( \begin{array}{c}}
\def\vece{\end{array} \right) }
\def\vectttb{\left( \begin{array}{c}}
\def\vecttte{\end{array} \right) }
\def\ref#1{\raisebox{.7ex}{\scriptsize #1)}}

\begin{document}

\bfr
\TEP{TEP-11R}

\TEP{June 1993}

\TEP{hep-th/9303049}
\efr
 \vs{1cm}

\begin{center}

{\Large
{\bf Exact Solutions to the Wheeler-DeWitt Equation of}}\vs{.7cm}

{\Large  {\bf Two Dimensional Dilaton Gravity}}\vs{2.0cm}

(revised)\vs{2.0cm}

{\bf Takayuki Hori}
\footnote[2]{E-mail address: e00353@sinet.ad.jp} \vs{0.7cm}

{\it Institute for Physics, Teikyo University }  \vs{0.4cm}

{\it Otsuka 359, Hachioji-shi, Tokyo 192-03, Japan } \vs{3.0cm}

{\bf Abstract}

\end{center}

\begin{normalsize}
\baselineskip=25pt

The two dimensional dilaton gravity with the cosmological term
and with an even number of matter fields minimally coupled to the
gravity is considered.
The exact solutions to the Wheeler-DeWitt equation  are obtained
in an explicit functional form, which contain an arbitrary
holomorphic function of the matter fields.

\newpage
\pagestyle{plain}

Recently the full quantum analysis of the two dimensional dilaton
gravity has been considered to be of great importance because it
must finally settle up the uncertainty on the  black hole
evaporation\ref{1} in two dimensions, the problem first posed by
Callan et al.\ref{2} in a semi-classical approximation.
There have been some attempts to analyse the model in the full
quantum setting (see, {\it e.g.}, ref.(3)).
Among them, however, the old-fashioned Wheeler-DeWitt (WD)
quantization scheme\ref{4} has scarcely been given attention\ref{5}.
This may be partially due to the conceptual trouble one encounters in
interpreting the wave function, when  one is forced to rely on the
asymptotic flatness to extract any measurable quantities.
Moreover the exact solutions are hardly found except in some mini
superspace models\ref{6}.

In this note we find the latter problem, {\it i.e.}, to find the
exact solutions to the WD equation is tractable in the two
dimensional dilaton gravity.
The problem on the interpretation of the wave function and the
analysis of the physical outcomes are postponed for a future study.

The classical action\ref{2} is
\bena
        I = \int \! \! d^2xe\{ e^{-2\f}[R(\w ) + k(\nabla \f )^2
+ 4\l ^2]    - \fr{1}{2}\sum_{a}(\nabla f_a)^2\} ,
\eena
where $\f$ and $f_a$ are the dilaton and matter fields,
respectively, and $\l$ is the cosmological constant.
$R(\w )$ is the curvature expressed in terms of the spin
connection ${\w _{\m a}}^{b}$, and $e$ is the determinant of
the zweibein, $e_{\m a}$.
The parameter $k$ depends on models and $k = 4$ corresponds
to ref.(2), where the classical equations of motion are simplified.
In the case $k = 8$ the WD equations have a different type of
solutions (see below).

The local Lorentz gauge is fixed by putting
\bena
              e_{\m a}  =  \mxxb \a & \b e^{-\r} \\
                                    0  &    e^{\r}
                           \mxxe  ,
\eena
where $\a$ and $\b$ are the lapse function and the shift
vector, respectively.
In ref.(7) the WD equations without the cosmological term were
obtained by the standard procedure\ref{4}.
For the case with general $k$ and with the cosmological term
the WD equations are expressed as
\bena
        \c \Ps  &\equiv& \left[ (\r '(x) - \dx )\fr{\d}{\d \r (x)}
+ \f '(x)\fr{\d}{\d \f (x)}
+ \sum_{a}f_a'(x)\fr{\d}{\d f_a(x)}\right] \Ps  = 0,\\
        \F  \Ps  &\equiv& \left[ \fr14 \fr{\d ^2}{\d \r (x)^2}
+ \fr{1}{k}\fr{\d ^2}{\d \r (x)\d \f (x)}
-  \fr{16}{k}e^{-4\f (x)}(\fr{8 - k}{4}\f '^2 - \f ^{(2)}
- \l ^2e^{2\r})\right. \nn
  & &  \siii \siii \left.
+ \fr{2}{k}e^{-2\f (x)}\sum_{a}\left( \fr{\d ^2}{\d f_a(x)^2}
- f'_a(x)^2\right) \right] \Ps  =  0,
\eena
where $x = x^1$ and primes denote the derivatives with
respect to $x$, and $\f ^{(2)} \equiv  \f '' - \r '\f '$
is the second order covariant derivative in the spatial
dimension.
In eq.(4) we should regularize the functional derivations
at the same points.
The appropriate rule is determined by the requirement that the
constraint algebra closes  or by the consistent condition for
the couple of the WD equations (3) and (4).
This requires $\fr{\d \vf (x)}{\d \vf (x)} = \d (x,x) = 0$
for $\vf  = (\r , \f , f_a)$ , and can be achieved by the
dimensional regularization\ref{4}.

To solve the WD equations let us assume the spatial manifold
is compact and there are two matter fields, $f_1$ and $f_2$.
We make the following ansatz for the wave functional $\Ps$
\def\zb{\bar{z}}
\bena
         \Ps  =
\exp{i\! \! \int \! \! dx[e^{-2\f}\f 'g(X, \f , z, \zb)
+ \fr{1}{4}(z\zb ' - \zb z')]},\\
          X \equiv  e^{-\r}\f ', \sii  z \equiv  f_1
+  if_2,
\eena
where $g$ is a function to be determined.
The $\c$-constraint, eq.(3), amounts to the general
coordinate invariance of the wave functional in the
spatial dimension.
Since $\f ', z'$ and $\zb '$ are scalar densities and
$X, \f , z$ and $\zb$ are scalars, $\Ps$ satisfies the
$\c$-constraint.
After some calculations we find the hamiltonian constraint,
eq.(4), can also be satisfied for a suitable $g$.
For $k \ne  8$ the solution is
\bena
      \Ps _{[A]} &=& \exp{i\int \! \! dx\left[
-e^{-2\f}\left( 2\f '^{-1}\f ^{(2)}\L + \fr{\bigl( e^{\fr{8
- k}{2}\f}A(z)\bigr) '}{X^2(1 + \L )}\right)
+ \fr{1}{4}(z\zb ' - \zb z')\right] },\\
          \L  &\equiv&  \sqrt{1 - \bigl( \fr{4}{8 - k}\l ^2
+ e^{\fr{8 - k}{2}\f}A(z)\bigr) X^{-2}},
\eena
where $A(z)$ is an  arbitrary {\it holomorphic} function of
$z \equiv  f_1 + if_2$.
For $k = 8$ we have a slightly different form of the solution:
\def\tilL{\tilde{\L}}
\bena
      \Ps _{[B]} &=&
\exp{i\int \! \! dx\left[ -e^{-2\f}\left( 2\f '^{-1}\f ^{(2)}\tilL
+ \fr{\left( B(z) - 2\l ^2\f \right) '}{X^2(1 + \tilL )}\right)
+ \fr{1}{4}(z\zb ' - \zb z')\right] }, \\
       \tilL  &\equiv&  \sqrt{1 - \bigl( B(z)
- 2\l ^2\f \bigr) X^{-2}},
\eena
where $B(z)$ is an arbitrary holomorphic function.
It is straightforward to generalize the solution to the models
with an even number of real matter fields.

If the cosmological term is absent there exists an another
type of solutions which cannot be expressed in the above
form with $\l  = 0$, {\it i.e.},
\bena
          \Ps _0  =
\exp{i\! \! \int \! \! dx[e^{-2\f}(-2\f '^{-1}\f ^{(2)}
+  M(\zb ,\f )')   +   \fr14 (z\zb ' - \zb z')]},
\eena
where $M$ is an arbitrary {\it anti}-holomorphic function
of $\zb$ and $\f$.
The solution found in ref.(7) is a special case of eq.(11).
Note that the space of the solutions expands at the point
$\l  = 0$.
Although we work within the ansatz (5) this implies some
sort of degeneracy at $\l  =0$.

Since the hamiltonian operator, $\F$, is real we see the
complex conjugates of the above solutions are other
independent solutions.
Thus we have the most general solution
\bena
         \Ps  =  \int \! {\cal D}A\, \left[ C_{[A]}\Ps _{[A]}
+ {\tilde{C}}_{[A^*]}{\Ps}^*_{[A^*]}\right] ,
\eena
where $\Ps _{[A]}$ is one of the above solutions, and
$C_{[A]}$ and $\tilde{C}_{[A^*]}$ are arbitrary
coefficients depending on the function form of $A$ and $A^*$.
If one proceeds to the third quantization the coefficients
$C(\tilde{C})$ are interpreted as annihilation (creation)
operators of the state specified by the particular configuration
corresponding to the holomorphic function $A(z)$.


Our solutions must contain much informations on the quantum states
of the two dimensional dilaton gravity, compared with those in
the mini-superspace approaches, since they  were obtained
without any approximations.
An investigation of the physical interpretation for the solution
is expected to give a final answer to the problem of the black
hole evaporation in two dimensions.\vss


I wish to thank H.Kakuhata and M.Kamata for discussions.
I am grateful to the members of High Energy Theory Group in
Tokyo Metropolitan University for kind hospitality.

\newpage

\baselineskip=20pt

\noi {\bf References}\vss

\noi 1) S. W. Hawking, {\it Commun. Math. Phys.},
{\bf 43}, (1975) 199.

\noi 2) C. Callan, S. B. Giddings, J. A. Harvey and A.
Strominger, {\it Phys. Rev.} {\bf D45},

\noi \sii  (1992) R1005.

\noi 3) S. Hirano, Y. Kazama and Y. Satoh, preprint
UT-Komaba 93-3, Mar. 1993.

\noi 4) B. S. DeWitt, {\it Phys. Rev.} {\bf 160} (1967) 1113;

\noi \sii  J. A. Wheeler, in {\it Relativity groups and
topology} (Gordon and Breach, New York, 1964).

\noi 5) A. Mikovi\'c, Imperial-TP/92-93/08, Nov. 1992.

\noi 6) U. H. Danielsson, CERN-TH. 6711/92, Nov. 1992.

\noi 7) T. Hori and M. Kamata, preprint TEP-10, Feb. 1993.

\newpage

\end{normalsize}
\end{document}